\documentclass{appolb}
\usepackage{graphicx}
\usepackage{nicefrac}
\usepackage{amsmath}
\usepackage[T1]{fontenc}
\usepackage[utf8]{inputenc}
\usepackage{BOONDOX-calo}

\usepackage{tikz}
\usepackage[compat=1.1.0]{tikz-feynman}
\tikzfeynmanset{every edge/.append style={very thick}}
\tikzset{
  cutline/.style={
    draw=orange,
    dashed,
    line width=4pt,
    dash pattern=on 10pt off 4pt   
  }
}
\usetikzlibrary{shadows}
\usetikzlibrary{shadows.blur}
\usetikzlibrary{backgrounds}
\usepackage{hyperref}
\hypersetup{
    colorlinks=true,
    urlcolor=blue,
    linkcolor = blue,
    urlcolor  = blue,
    citecolor = blue,
    anchorcolor = blue,
    pdftitle={TFF-Roper},
    pdfauthor={friends},
    pdfsubject={Roper 90}}
\begin{document}
\title{Comments on Baryon Transition Form Factors%
\thanks{Prepared for the occasion of David Roper's 90$^{\rm th}$ birthday.}%
}
\author{Christoph Hanhart
\address{Institute for Advanced Simulation (IAS-4), Forschungszentrum J\"ulich, \\ J\"ulich, Germany}
\\[3mm]
{Maxim Mai 
\address{Albert Einstein Center for Fundamental Physics, Institute for Theoretical Physics, University of Bern, Sidlerstrasse 5, 3012 Bern, Switzerland \& The George Washington University, Washington, DC 20052, USA}
}
\\[3mm]
{Ulf-G. Mei{\ss}ner
\address{Helmholtz-Institut f\"{u}r Strahlen- und Kernphysik and Bethe Center for Theoretical Physics, Universit\"{a}t Bonn, D-53115 Bonn, Germany \&
Institute for Advanced Simulation (IAS-4), Forschungszentrum J\"ulich, J\"ulich, Germany \&
Peng Huanwu Collaborative Center for Research and Education, International Institute for Interdisciplinary and Frontiers, Beihang University, Beijing 100191, China}
}
\\[3mm]
Deborah R\"onchen
\address{Institute for Advanced Simulation (IAS-4), Forschungszentrum J\"ulich, \\ J\"ulich, Germany}
}
\maketitle
\begin{abstract}
We discuss in rather general terms the properties of space-like baryon transition form factors. In particular, we argue why these are necessarily complex-valued, what can be deduced from the respective phase motion and why dealing with real valued transition form factors in general leads to misleading results. For illustration the transition form factors for the Roper resonance as derived in the J\"ulich-Bonn-Washington framework are discussed. 
\end{abstract}

\pagebreak

\section{Introduction}
\label{sec:intro}

Physical states can appear either as bound states, virtual states or resonances. Bound states are stable systems, with normalized wave functions. Mathematically, they manifest themselves as poles of the $S$-matrix on the physical Riemann sheet. Examples for those poles, allowed only below the lowest threshold of the system, are the proton, the neutron (although the neutron can decay weakly, its life time is so long that it often can be treated as stable) and the deuteron as a bound system of proton and neutron. Virtual states and resonances are also connected to poles of the $S$-matrix, however,  those are located on an unphysical Riemann sheet and their wave functions are not normalizable. The poles for virtual states are located on the real axis below the lowest threshold, those of resonances in the complex plane. Typically, both kinds of poles leave an imprint on observables. A famous virtual state is the neutron-neutron state, located only about 90~keV below threshold on the second sheet.  Examples for resonances are the $\rho(770)$ meson and the $\Delta(1232)$ baryon---and the Roper resonance, $N^*(1440)$~\cite{Roper:1964zza}.

A key question in hadron spectroscopy is the composition of states. The most simple realization of the quark model assigns baryons as three-quark states and mesons as quark--anti-quark systems. However, in recent years it became evident that QCD generates a lot more complicated structures, especially multi-quarks, where the quark content exceeds the numbers given above for the naive quark model. In the doubly heavy quark sector the number of candidates for multi-quark states now exceeds that for regular quarkonia as soon as the energy is above the lowest open-flavor threshold; see Refs.~\cite{Hosaka:2016pey, Esposito:2016noz, Guo:2017jvc, Olsen:2017bmm, Karliner:2017qhf, Brambilla:2019esw, Yang:2020atz, Chen:2022asf, Meng:2022ozq} for recent reviews. But also in the light quark sector with, e.g., the $\Lambda(1405)$ there is a well established state that does not fit into a conventional classification~\cite{Kaiser:1995eg, Oller:2000fj}, it even reveals a two-pole structure. For dedicated reviews see, e.g., \cite{Mai:2020ltx, Hyodo:2020czb}. And also the Roper resonance shows unusual properties, see, e.g.,~\cite{Krehl:1999km}. For example it is lighter than the first negative parity excitation of the nucleon, the $S_{11}(1535)$, and the $\pi N$ inelasticity drops very steeply right in the mass range of the Roper state. It is also the first baryon resonance which has a sizable decay rate into a three-particle final-state with two pions, making it an important testbed for three-particle dynamics. 
Through this, the Roper resonance became also a long-term goal motivating a lot of ab-initio lattice QCD efforts in conducting calculations and improving the theoretical tool-box mapping finite-volume results to infinite-volume quantities~\cite{Hansen:2014eka, Mai:2017bge, Severt:2022jtg, Hansen:2025oag,Sharpe:2026mtt}. Notable intermediate steps in this regard have recently been first-ever calculations of 3-body resonances $\omega(782)$~\cite{Yan:2024gwp} and the first excited state of pion $\pi(1300)$~\cite{Yan:2025mdm}---a state 10 times heavier than its ground state, and nearly as heavy as the Roper resonance itself.

In resolving the structure and composition of resonant states, photons (real or virtual) provide a crucial scanning probe. Indeed, electro-magnetic transition form factors, to be introduced in section~\ref{sec:transitionformfactors}, are believed to provide better insight into the structure of the exotic candidates. In the next section we will first provide a quick review of the properties of resonances. For more details we refer to the review "Resonances" in the Review of Particle Physics (RPP)~\cite{ParticleDataGroup:2024cfk} and references therein as well as Refs.~\cite{Mai:2022eur, Willenbrock:2025otp}.

\section{On the Properties of Resonances}
\label{sec:properties-of-resonances}

A resonance is characterized by its pole locations and its residues. Here we wrote deliberately plural for both properties, since a resonance has poles on various Riemann sheets and, when it couples to various channels, also various residues (which are different for the different poles). However, the significance of a given resonance pole depends on its distance to the physical axis. Thus in many cases the observables are mostly influenced by a single pole and it is this single pole that is quoted in the RPP. For simplicity this is the case we focus on from here on. 

Let us assume a scattering amplitude contains a pole. Then it is always possible to split it according to
\begin{equation}
    T(s)_{ij}=T(s)_{{\rm bg}\, ij}+T(s)_{{\rm R}\, ij}\,,
    \label{poledecomp}
\end{equation}
where the first term, $T(s)_{\rm bg}$ is regular at $s=s_{\rm R}$, and the second term, $T(s)_{\rm R}$ has a pole at that location. The indices $i$ and $j$ specify the pertinent channels---in case of the Roper resonance discussed below those are $\pi N$ and $\pi\pi N$ (typically parameterized as $\rho N$, $\sigma N$ and $\pi \Delta$). The pole  location $s_{\rm R}$ should in principle carry a label of the sheet where the pole is located, which we omit here to simplify notations.

Note that the decomposition provided in Eq.~(\ref{poledecomp}) is not unique. Especially for real values of $s$ one can straightforwardly shift strength from one term to the other. The only way to define resonance properties unambiguously and independently of a particular reaction is through an analytic continuation of the physical amplitude to the resonance pole $s_{\rm R}$ and then to extract the pole residues.

\begin{figure}
\centering
\begin{tikzpicture}[baseline=(c.base)]
\begin{feynman}
    \vertex (piout) at (-1,-1);
    \vertex (Nout)  at (-1,+1);
    \vertex (a) at (0,0);
    \vertex (Rin) at (+1,0);
    \diagram*{
    (Nout) -- [scalar] (a) -- [double] (Rin),
    (piout)-- [plain] (a)
    };
    \vertex[draw, fill=orange, circle, minimum size=8mm, inner sep=0pt] (c) at (0,0) {\( \Gamma \)};
\end{feynman}
\end{tikzpicture}
=
\begin{tikzpicture}[baseline=(c.base)]
\begin{feynman}
    \vertex (piout) at (-1,-1);
    \vertex (Nout)  at (-1,+1);
    \vertex (a) at (0,0);
    \vertex (Rin) at (+1,0);
    \diagram*{
      (Nout) -- [scalar] (a) -- [double] (Rin),
      (piout)-- [plain] (a)
    };
    \vertex[draw, fill=black, circle, minimum size=3mm, inner sep=0pt] (c) at (0,0) {\( \Gamma \)};
\end{feynman}
\end{tikzpicture}
+
\begin{tikzpicture}[baseline=(c.base)]
\begin{feynman}
    \vertex (piout) at (-2,-1);
    \vertex (Nout)  at (-2,+1);
    \vertex (a) at (0,0);
    \vertex (a1) at (-1,+0.5);
    \vertex (a2) at (-1,-0.5);
    \vertex (Rin) at (+1,0);
    \diagram*{
    (Nout) -- [scalar] (a1) -- [scalar, bend left=40] (a) -- [double] (Rin),
    (piout)-- [plain] (a2) -- [plain, bend left=-40] (a)
    };
    \vertex[draw, fill=gray!30, ellipse, 
        minimum width=10mm,
        minimum height=15mm] (c) at (-1,0) {\( T_{\rm bg}\)};
    \vertex[draw, fill=black, circle, minimum size=3mm, inner sep=0pt] (c) at (0,0) {\( \Gamma \)};
\end{feynman}
\end{tikzpicture}
\caption{Diagrammatic representation of the vertex function $\pi N\to R$, denoting nucleon/pion/resonance states by a full/dashed/double line, respectively. The full vertex function ($\Gamma$) is given by the sum of the bare vertex ($\bullet$) plus a contribution where the external particles interact via the background interaction ($T_{\rm bg}$), shown as the gray ellipse.}
\label{fig:vertexfunction}
\end{figure}

In this section we follow Ref.~\cite{Heuser:2024biq}
and discuss a special form of the decomposition 
given in Eq.~(\ref{poledecomp}). In particular, we assume that the background is diagonal in the channel-space, although in general this is not the case, and constructed in a unitary way. Since the full $T$-matrix is unitary as well, $T_{\rm R}$ cannot be unitary individually. Instead it takes the form (although we discuss predominantely baryons in this note we refrain from keeping track with the Dirac structure to simplify notations)~\cite{Nakano:1982bc,Hanhart:2012wi} 
\begin{equation}
T_{{\rm R}\, ij} = -
\Gamma(s)_{{\rm out}\, k}^\dagger \delta_{ki} \frac{g_i n_i(s) n_j(s) g_j}{s-m^2-\sum_a g_a^2\Sigma(s)_a}\delta_{jl}\Gamma(s)_{{\rm in}\, l} \,,
\label{eq:TRdetail}
\end{equation}
with
\begin{eqnarray} 
\mbox{Disc}\left[\Gamma(s)_{{\rm out}\ k}^\dagger\right]&=&2i \ T(s)_{{\rm bg}\,kk}^*\, \rho_k(s)\,  \Gamma(s)_{{\rm out}\, k}^\dagger\ , \\
\mbox{Disc}\left[\Gamma(s)_{{\rm in}\ k}\right]&=&2i \
\Gamma(s)_{{\rm in}\, k}\, \rho_k(s)\, T(s)_{{\rm bg}\, kk}^* \ , \\
\mbox{Disc}\left[\Sigma(s)_k\right]&=& 2i\ \Gamma(s)_{{\rm in}\, k}\rho_k(s) n_k(s)^2\Gamma(s)^\dagger_{{\rm out}\, k}  \ ,
\end{eqnarray}
where $\rho_k(s)=q_k/(8\pi \sqrt{s})$ and $n_k(s)=(q_k/q_0)^{\ell_k}F(q_k/q_0)_{\ell_k}$ denote the phase-space factor and the centrifugal barrier factor with respect to the channel specific angular momentum $\ell_k$, respectively. The function $F_{\ell}$ denote phenomenological form factors necessary to tame the otherwise unlimited growth of the centrifugal barrier term for $\ell_k>0$. One typically chooses~\cite{Blatt:1952ije,VonHippel:1972fg,Chung:1995dx}
\begin{equation}
    F(z)_0^2=1\ , \ \ F(z)_1^2=(1+z^2)^{-1} \ , \ \ F(z)_2^2=(9+3z^2+z^4)^{-1} \,, ... \,.
\end{equation}

A minimal resonance model, with $T_{\rm bg}=0$, in a single channel reaction is characterized by two parameters, the bare mass and the bare 
coupling\footnote{Note that these bare quantities are devoid of any physical meaning.}. Those can be employed to exactly reproduce the real and imaginary part the most significant pole location of a resonance. In this way the residue is fixed automatically as well. In Ref.~\cite{Heuser:2024biq} it is demonstrated that in this way the residue of the $\rho$ meson is described well. In contrast to this, a proper description of the $f_0(500)$ residue was possible only, when a background was included. Interestingly, this background, when adjusted to reproduce both the absolute value of the residue and its phase, automatically introduced an Adler zero to the scattering amplitude as demanded by chiral symmetry, nicely showing the intimate relation between the properties of the $f_0(500)$ and chiral symmetry. The way, how the background modifies the vertex function is illustrated in Fig.~\ref{fig:vertexfunction}. In particular, there are phases from both the background $T$-matrix as well as the intermediate two-particle state.

The above mentioned differences between the $\rho$ and the $\sigma$ amplitude can be interpreted such that the $\rho$ has a conventional $\bar qq$ structure while the $f_0(500)$ owes its existence a non-perturbative $\pi\pi$ interactions---this conclusion is in line with other, independent studies; see Ref.~\cite{Pelaez:2015qba} for a review. This important information is encoded in both the pole residue and the phase. It is obvious that also the transition form factor is sensitive to the internal structure of a given state. Even more so, since the virtuality of the photon provides an additional degree of freedom. This is discussed in the next section.

\section{Transition form factors}
\label{sec:transitionformfactors}

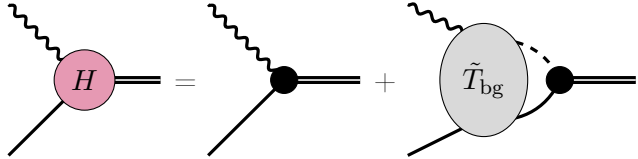
\begin{figure}
\centering
\begin{tikzpicture}[baseline=(c.base)]
\begin{feynman}
    \vertex (piout) at (-1,-1);
    \vertex (Nout)  at (-1,+1);
    \vertex (a) at (0,0);
    \vertex (Rin) at (+1,0);
    \diagram*{
    (Nout) -- [photon] (a) -- [double] (Rin),
    (piout)-- [plain] (a)
    };
    \vertex[draw, fill=purple!40, circle, minimum size=8mm, inner sep=0pt] (c) at (0,0) {\( H \)};
\end{feynman}
\end{tikzpicture}
=
\begin{tikzpicture}[baseline=(c.base)]
\begin{feynman}
    \vertex (piout) at (-1,-1);
    \vertex (Nout)  at (-1,+1);
    \vertex (a) at (0,0);
    \vertex (Rin) at (+1,0);
    \diagram*{
      (Nout) -- [photon] (a) -- [double] (Rin),
      (piout)-- [plain] (a)
    };
    \vertex[draw, fill=black, circle, minimum size=3mm, inner sep=0pt] (c) at (0,0) {\( \Gamma \)};
\end{feynman}
\end{tikzpicture}
+
\begin{tikzpicture}[baseline=(c.base)]
\begin{feynman}
    \vertex (piout) at (-2,-1);
    \vertex (Nout)  at (-2,+1);
    \vertex (a1) at (-1,+0.5);
    \vertex (a2) at (-1,-0.5);
    \vertex (a) at (0,0);
    \vertex (Rin) at (+1,0);
    \diagram*{
    (Nout) -- [photon] (a1) -- [scalar, bend left=40] (a) -- [double] (Rin),
    (piout)-- [plain] (a2) -- [plain, bend left=-40] (a)
    };
    \vertex[draw, fill=gray!30, ellipse, 
        minimum width=10mm,
        minimum height=15mm] (c) at (-1,0) {\( \tilde T_{\rm bg}\)};
    \vertex[draw, fill=black, circle, minimum size=3mm, inner sep=0pt] (c) at (0,0) {\( \Gamma \)};
\end{feynman}
\end{tikzpicture}
\caption{Diagrammatic representation of the transition form factor $H$, denoting    nucleon/photon/resonance states by a full/wavy/double line, respectively.
  The full form factor is given by the sum of the bare vertex ($\bullet$) plus a contribution where the external particles interact via the background interaction $\tilde T_{\rm bg}$, shown as the gray ellipse.}
\label{fig:formfactor}
\end{figure}

As discussed in the previous section also for transition from factors  (TFFs) only resonance properties at the pole are well defined and accordingly also transition form factors should be extracted at the resonance pole. To find the proper theoretical expressions, the vertex function $\Gamma_{\rm in}$, that appears in  $T_{\rm R}$, provided in Eq.~\eqref{eq:TRdetail}, needs to be gauged. For a generic procedure in the context of Effective Field Theories see, e.g., Refs.~\cite{vanAntwerpen:1994vh, Borasoy:2007ku, Ruic:2011wf,Mai:2012wy,Bruns:2020lyb}. In this way the information about the presence or absence of a background term gets transferred also to the transition form factor. In particular, the transition form factor is necessarily complex valued, although the photon is typically space like, such that crucial information is encoded in the emerging phase: Since the two-particle intermediate state shown explicitly in the diagram on the far right of Fig.~\ref{fig:formfactor} can typically go on-shell for a resonance, which has open decay channels, there is an unavoidable non-trivial phase motion for all those resonances, where the presence of background terms are crucial (see discussion in the previous section) even at the pole.

It is important to stress that, since pole and non-pole contributions cannot be separated model-independently as stated above, it is also not possible to model-independently disentangle the so-called meson cloud contributions to the transition form factor and pure resonance contributions, see also the discussion in~\cite{Bernard:1998gv}. However, it follows from the discussion in the previous paragraph, that if the phases of the hadronic residues differ significantly from those of the transition form factors there must be background contributions present pointing at a non-trivial structure of the resonance. 

\bigskip

The electromagnetic TFFs of a stable state (e.g., $N$) to an excited state (e.g., $N^*(1440)$) cannot be measured directly, since the resonance is not an asymptotically stable state but decays for example into $\pi N$ or $\pi\pi N$ final states. TFFs, therefore, need to be extracted from electroproduction amplitudes, e.g., for $\gamma^*N\to \pi N$, constrained by experimental data from, e.g., CLAS@JLAB~\cite{CLAS:2002xbv, CLAS:2009ces, CLAS:2012wxw,  Mokeev:2015lda}. In the following, we show a possible procedure for the extraction of TFFs from experimental electroproduction data. 

\begin{figure}[t]
    \centering
    \includegraphics[width=0.5\linewidth]{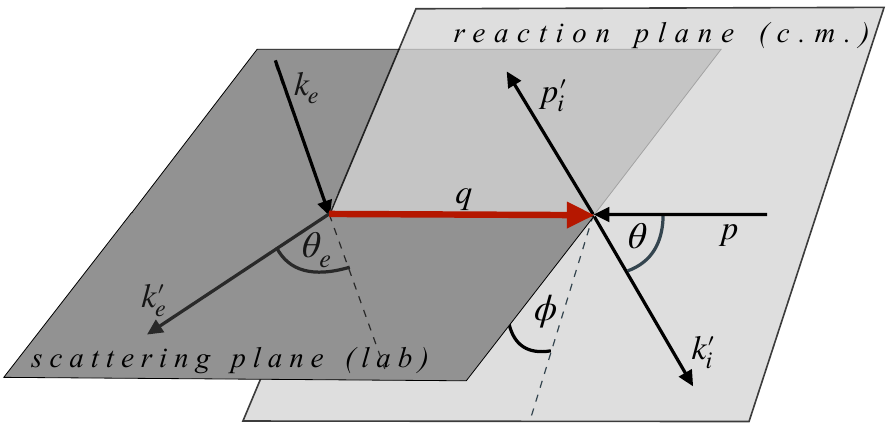}\\
    \rule{\textwidth}{0.5pt}\\[1mm]
    \includegraphics[width=\linewidth]{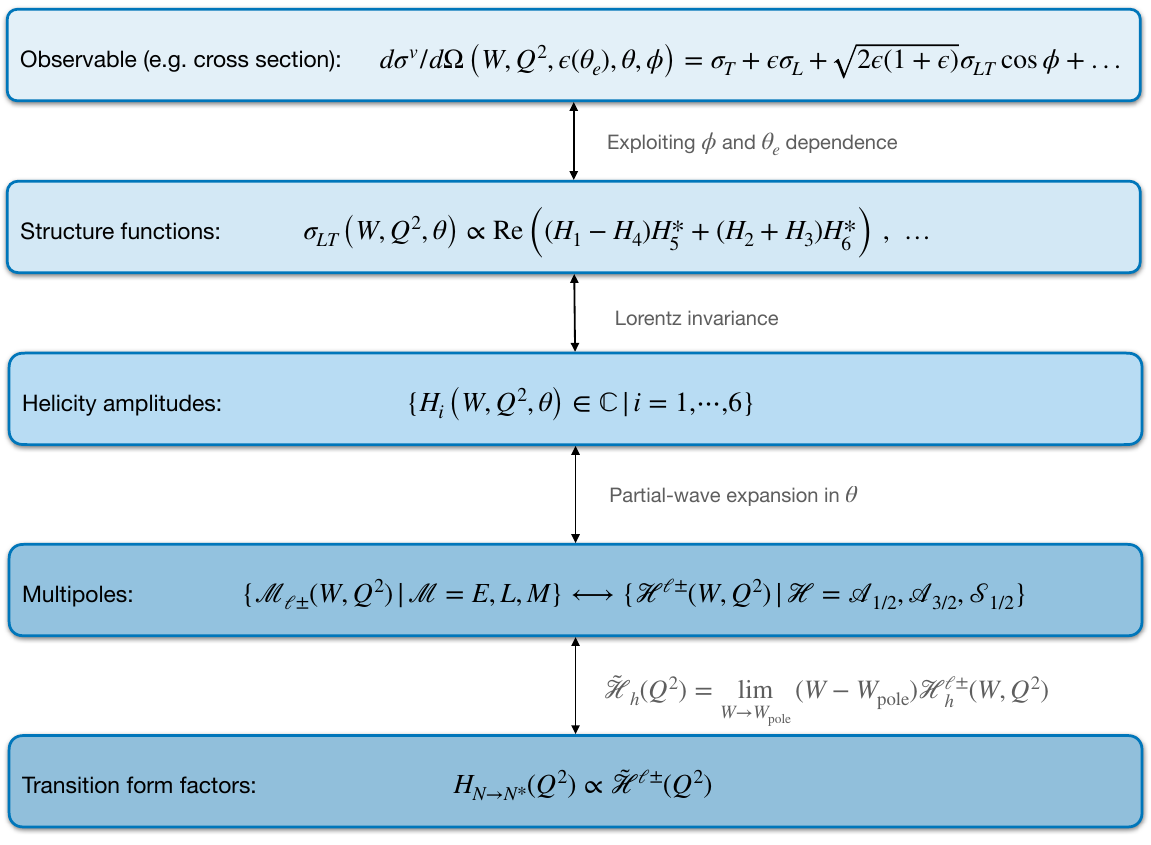}
    \caption{Top panel: Degrees of freedom of a one-meson electroproduction reaction. Bottom panel: Connection between reaction independent transition form factors $H_{N\to N^*}$ and observables. For more details and formulas see Refs.~\cite{Mai:2021vsw,Mai:2021aui,Mai:2023cbp}.}
    \label{FIG:kinematics-observables}
\end{figure}

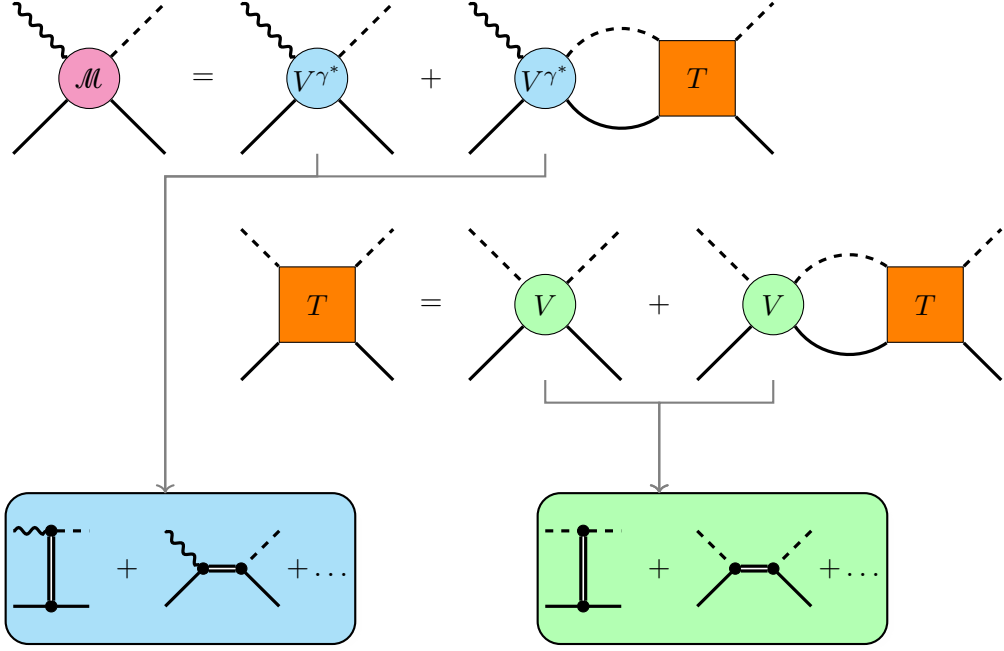
\begin{figure}
    \centering
    \begin{tikzpicture}
  \draw[thick,fill=cyan!30,rounded corners=3mm,blur shadow={shadow opacity=0.4}] (2.9,-10.5) rectangle (7.5,-8.5);
  \draw[thick,fill=green!30,rounded corners=3mm,blur shadow={shadow opacity=0.4}] (9.9,-10.5) rectangle (14.5,-8.5);

  \begin{feynman}
  \vertex (Nv2f) at (3,-2);
  \vertex (f6hY) at (3,-4);
  \vertex[draw, fill=magenta!50, circle, minimum size=8mm, inner sep=0pt] (3fdg) at (4,-3) {$\mathcal{M}$};
  \vertex (AttQ) at (5,-2);
  \vertex (Xbrx) at (5,-4);
  \vertex (8ZPc) at (5.5,-3) {$=$};
  \vertex (R8h5) at (6,-2);
  \vertex (Jm97) at (6,-4);
  \vertex[draw, fill=cyan!30, circle, minimum size=8mm, inner sep=0pt] (HgtQ) at (7,-3) {$V^{\gamma^*}$};
  \vertex (Bf74) at (8,-2);
  \vertex (1nbR) at (8,-4);
  \vertex (K9Cu) at (8.5,-3) {$+$};
  \vertex (p1Zd) at (9,-2);
  \vertex (gkeU) at (9,-4);
  \vertex[draw, fill=cyan!30, circle, minimum size=8mm, inner sep=0pt] (WEBt) at (10,-3) {$V^{\gamma^*}$};
  \node[draw, fill=orange, rectangle, minimum size=10mm, inner sep=2pt] (SIWR) at (12,-3) {$T$};
  \vertex (QoYP) at (13,-2);
  \vertex (Y1R7) at (13,-4);
  \diagram*{
  (f6hY) -- [plain] (3fdg) -- [plain] (Xbrx),
  (Nv2f) -- [photon] (3fdg) -- [scalar] (AttQ),
  (Jm97) -- [plain] (HgtQ) -- [plain] (1nbR),
  (R8h5) -- [photon] (HgtQ) -- [scalar] (Bf74),
  (gkeU) -- [plain] (WEBt) -- [plain, bend left=-45] (SIWR) -- [plain] (Y1R7),
  (p1Zd) -- [photon] (WEBt) -- [scalar, bend left=45] (SIWR) -- [scalar] (QoYP),
  };

  \node[draw, fill=orange, rectangle, minimum size=10mm, inner sep=2pt] (SIWR) at (7,-6) {$T$};  
  \vertex (d1) at (6,-7);
  \vertex (d2) at (6,-5);
  \vertex (d3) at (8,-5);
  \vertex (d4) at (8,-7);
  \diagram*{
  (d1) -- [plain] (SIWR) -- [plain] (d4),
  (d2) -- [scalar] (SIWR) -- [scalar] (d3)
  };
  \node at (8.5,-6) {$=$};
  \vertex (4Qik) at (9,-5);
  \vertex (5RAl) at (9,-7);
  \vertex[draw, fill=green!30, circle, minimum size=8mm, inner sep=0pt] (KGJ0) at (10,-6) {$V$};
  \vertex (aRWG) at (11,-5);
  \vertex (rbnx) at (11,-7);
  \node at (11.5,-6) {$+$};
  \vertex (m2r4) at (12,-5);
  \vertex (QwYl) at (12,-7);
  \vertex[draw, fill=green!30, circle, minimum size=8mm, inner sep=0pt] (SwXJ) at (13,-6) {$V$};
  \node[draw, fill=orange, rectangle, minimum size=10mm, inner sep=2pt] (ikp9) at (15,-6) {$T$};
  \vertex (HOwp) at (16,-5);
  \vertex (L0Fm) at (16,-7);
  \diagram*{
  (5RAl) -- [plain] (KGJ0) -- [plain] (rbnx),
  (4Qik) -- [scalar] (KGJ0) -- [scalar] (aRWG),
  (QwYl) -- [plain] (SwXJ) -- [plain, bend left=-45] (ikp9) -- [plain] (L0Fm),
  (m2r4) -- [scalar] (SwXJ) -- [scalar, bend left=45] (ikp9) -- [scalar] (HOwp),
  };

  \vertex (d1) at (3,-9);
  \vertex (d2) at (3,-10);
  \node[dot] (d3) at (3.5,-9);
  \node[dot] (d4) at (3.5,-10) ;
  \vertex (d5) at (4,-9);
  \vertex (d6) at (4,-10);
  \node at (4.5,-9.5) {$+$};
  \diagram*{  
  (d1) -- [photon] (d3) -- [scalar] (d5),
  (d2) -- [plain] (d4) -- [plain] (d6),
  (d3) -- [double] (d4)
  };
  \vertex (d1) at (5,-9);
  \vertex (d2) at (5,-10);
  \node[dot] (d3) at (5.5,-9.5);
  \node[dot] (d4) at (6,-9.5);
  \vertex (d5) at (6.5,-9);
  \vertex (d6) at (6.5,-10);
  \node at (7,-9.5) {$+\ldots$};
  \diagram*{  
  (d1) -- [photon] (d3) -- [plain] (d2),
  (d5) -- [scalar] (d4) -- [plain] (d6),
  (d3) -- [double] (d4)
  };

  \vertex (d1) at (10,-9);
  \vertex (d2) at (10,-10);
  \node[dot] (d3) at (10.5,-9);
  \node[dot] (d4) at (10.5,-10);
  \vertex (d5) at (11,-9);
  \vertex (d6) at (11,-10);
  \node at (11.5,-9.5) {$+$};
  \diagram*{  
  (d1) -- [scalar] (d3) -- [scalar] (d5),
  (d2) -- [plain] (d4) -- [plain] (d6),
  (d3) -- [double] (d4)
  };
  \vertex (d1) at (12,-9);
  \vertex (d2) at (12,-10);
  \node[dot] (d3) [dot] at (12.5,-9.5);
  \node[dot] (d4) [dot] at (13,-9.5);
  \vertex (d5) at (13.5,-9);
  \vertex (d6) at (13.5,-10);
  \node at (14,-9.5) {$+\ldots$};
  \diagram*{  
  (d1) -- [scalar] (d3) -- [plain] (d2),
  (d5) -- [scalar] (d4) -- [plain] (d6),
  (d3) -- [double] (d4)
  };
  \draw[->, thick,gray] (7,-4)  to (7,-4.3) to (5,-4.3) to (5,-8.5);
  \draw[thick,gray] (10,-4)  to (10,-4.3) to (5,-4.3) to (5,-8.5);
  \draw[->, thick,gray] (10,-7) to (10,-7.3) to (11.5,-7.3) to (11.5,-8.5);
  \draw[->, thick,gray] (13,-7) to (13,-7.3) to (11.5,-7.3) to (11.5,-8.5);
  \end{feynman}
\end{tikzpicture}
    \caption{JBW multipole parametrization. For any fixed quantum number, the multiples $\{\mathcal{M}_{\ell\pm}(W,Q^2)| \mathcal{M}=E,M,L\}$ are determined through a set of coupled-channel integral equations with respect to the scattering potential $V$ and the electroproduction term $V^{\gamma^*}$.}
    \label{fig:JBW-method}
\end{figure}

First, we note that any electroproduction observable to a two-body final state is a function of five independent kinematic variables. One useful choice of those is depicted in Fig.~\ref{FIG:kinematics-observables}. The transition of  successively separating off the angular dependence is visualized in the bottom part of Fig.~\ref{FIG:kinematics-observables}, leaving one with electric, magnetic and longitudinal multipoles $\{\mathcal{M}_{\ell\pm}(W,Q^2)| \mathcal{M}=E,M,L\}$. Here, $W=\sqrt{s}$, $Q^2:=-q^2$ and $\ell$ denote total energy, photon virtuality and the relative angular momentum of the pion-nucleon pair, respectively. The total angular momentum $J=\ell\pm 1/2$ is specified via $\ell\pm$  (for final meson states with non-vanishing spin the notation needs to be generalized). Different coupled-channel models (e.g., MAID~\cite{Tiator:2018pjq}, ANL/Osaka~\cite{Kamano:2018sfb}, JBW~\cite{Mai:2021vsw}, see also recent dedicated review~\cite{Doring:2025sgb}) can be compared conveniently on the level of these in general complex valued multipoles. For example, in the JBW approach (see Fig.~\ref{fig:JBW-method} for a visualization), 
which we take here as an example, the multipoles are parametrized as
\begin{align}
    \mathcal{M}_{\ell\pm}&=
    V^{\gamma^*}_{\ell\pm}
    +\int\limits_0^\infty dp\, p^2\, T_{\ell\pm}^{\phantom{\gamma^*}}GV^{\gamma^*}_{\ell\pm}
    \,,
    \label{eq:M}
\end{align}
where we have suppressed channel and isospin indices of the coupled-channel problem as well as all kinematic variables to ease the notation. Note also that an additional factorization of $V^{\gamma^*}_{\ell \pm}$ has been undertaken in the original JBW-formulation. A complete set of formulas can be found in Refs.~\cite{Mai:2021vsw, Mai:2021aui, Mai:2023cbp}. In the coupled-channel space, the elements denote:
\begin{itemize}
    \item $V^{\gamma^*}$: photon induced meson-baryon production potential; 
    \item $T_{\ell\pm}$: coupled-channel meson-baryon scattering amplitude (contrained/fixed in purely hadronic reactions);
    \item $G$: meson-baryon Green's function (fixed by the formalism); 
\end{itemize}
The meson-baryon scattering amplitude is itself a solution of a Lippmann-Schwinger coupled-channel equation
\begin{align}
    T_{\ell \pm}&=
    V_{\ell \pm}+\int\limits_0^\infty dp\, p^2\, V_{\ell \pm} G T_{\ell \pm}\,.
    \label{eq:T}
\end{align}
The scattering potential $V$ is derived from a chiral-symmetric Lagrangian and includes $s$-channel processes accounting for genuine resonances, as well as $t$- and $u$-channel exchanges of mesons and baryons, respectively, and contact diagrams. This separation is depicted in Fig.~\ref{fig:JBW-method}. Note also that certain resonances are dynamically generated in this approach. Accordingly, the photon-vertex separates as (again suppressing most of the dependencies on kinematic variables)
\begin{align}
    V^{\gamma^*}_{\ell \pm}&=
    F^{NP}(Q^2)\alpha^{NP}+F^{P}(Q^2)\frac{\gamma_{R\to MB}\gamma_{\gamma^*N\to R}}{W-m_b}
    \,,
    \label{eq:V}
\end{align}
where the resonance bare mass ($m_b$) and its bare couplings to the pertinent outgoing meson-baryon pair ($\gamma_{R\to MB}$)  are fixed from the same Lagrangian as the hadronic scattering potential shown in Eq.~(\ref{eq:T}), while the bare couplings to the incoming $\gamma N$ ($\gamma_{\gamma^*N\to R}$) are parameterized by polynomials. The same applies to the quantity $\alpha^{NP}$, that simulates the interaction of the photon with the background. The channel-(in)dependent form-factors $F^{NP}(F^{P})$ parametrize the sole photon-virtuality dependence of the potential. Constraints ensuring Siegert's condition are implemented, while the approach also respects final-state unitarity and gauge invariance by construction. 

The above procedure contains a set of free parameters determined from fits to the experimental data on elastic ($\pi N\to \pi N$) and
inelastic ($\pi N\to \eta N,K\Lambda,K\Sigma$) $\pi N$ reactions~\cite{Ronchen:2012eg} as well as various reactions induced by real ($\gamma p \to \pi N,\eta N,K\Lambda$~\cite{Ronchen:2018ury}) and virtual photons ($\gamma^* p \to \pi N,\eta N,K\Lambda$~\cite{Mai:2021vsw, Mai:2021aui, Mai:2023cbp} ) counting overall 8k+40k+110k=158k data. The obtained multipoles $\mathcal{M}_{\ell \pm}$ can be accessed through the dedicated JBW-homepage [\url{https://jbw.phys.gwu.edu}]. 

\begin{figure}[t]
    \centering
    \includegraphics[width=0.47\linewidth]{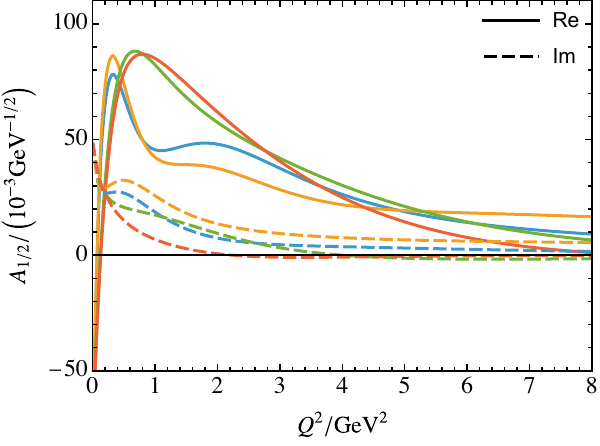}
    \includegraphics[width=0.49\linewidth]{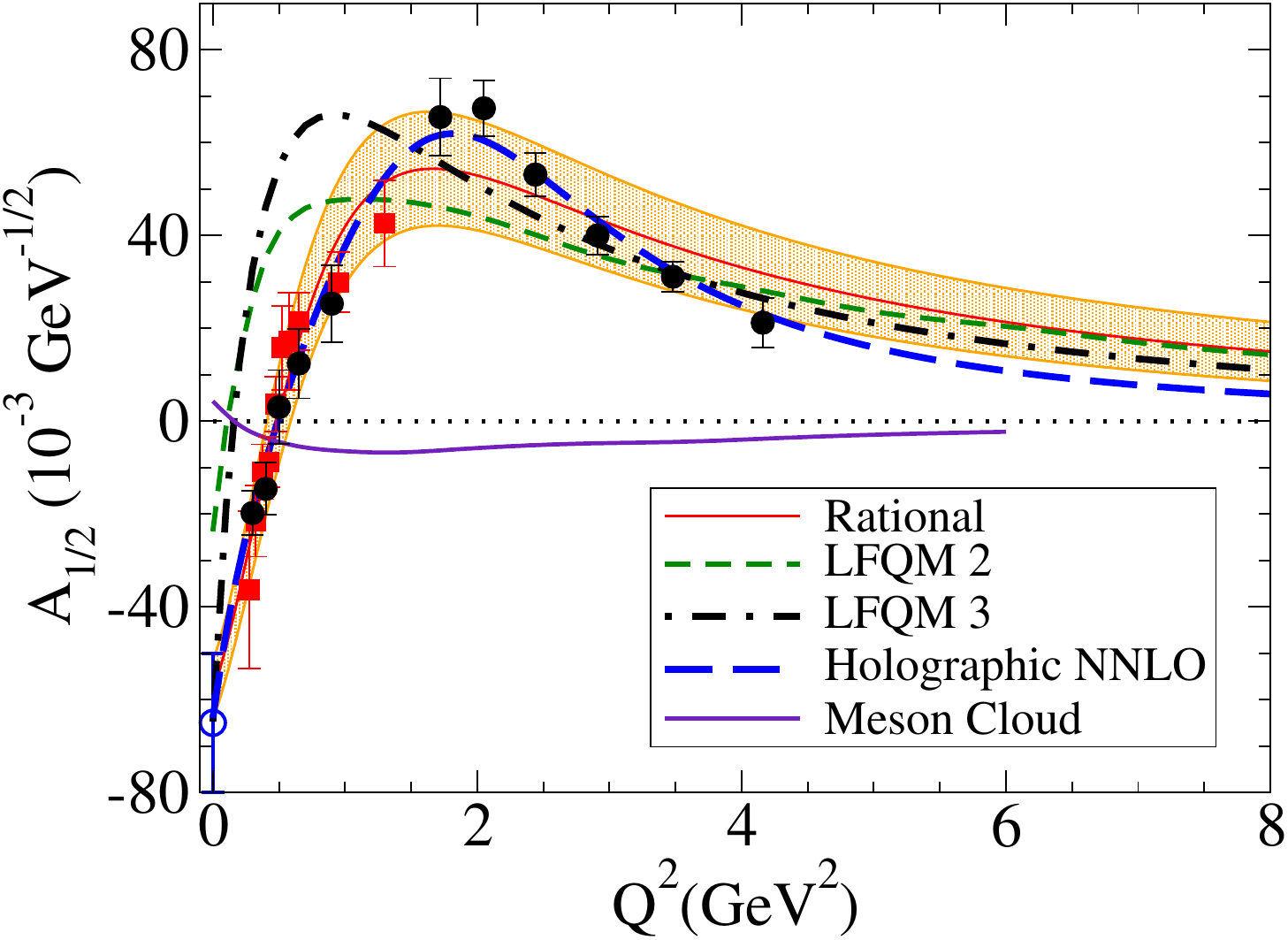}
    \includegraphics[width=0.47\linewidth]{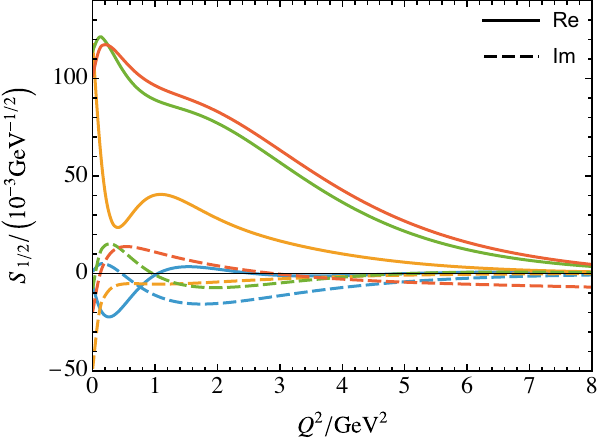}
    \includegraphics[width=0.49\linewidth]{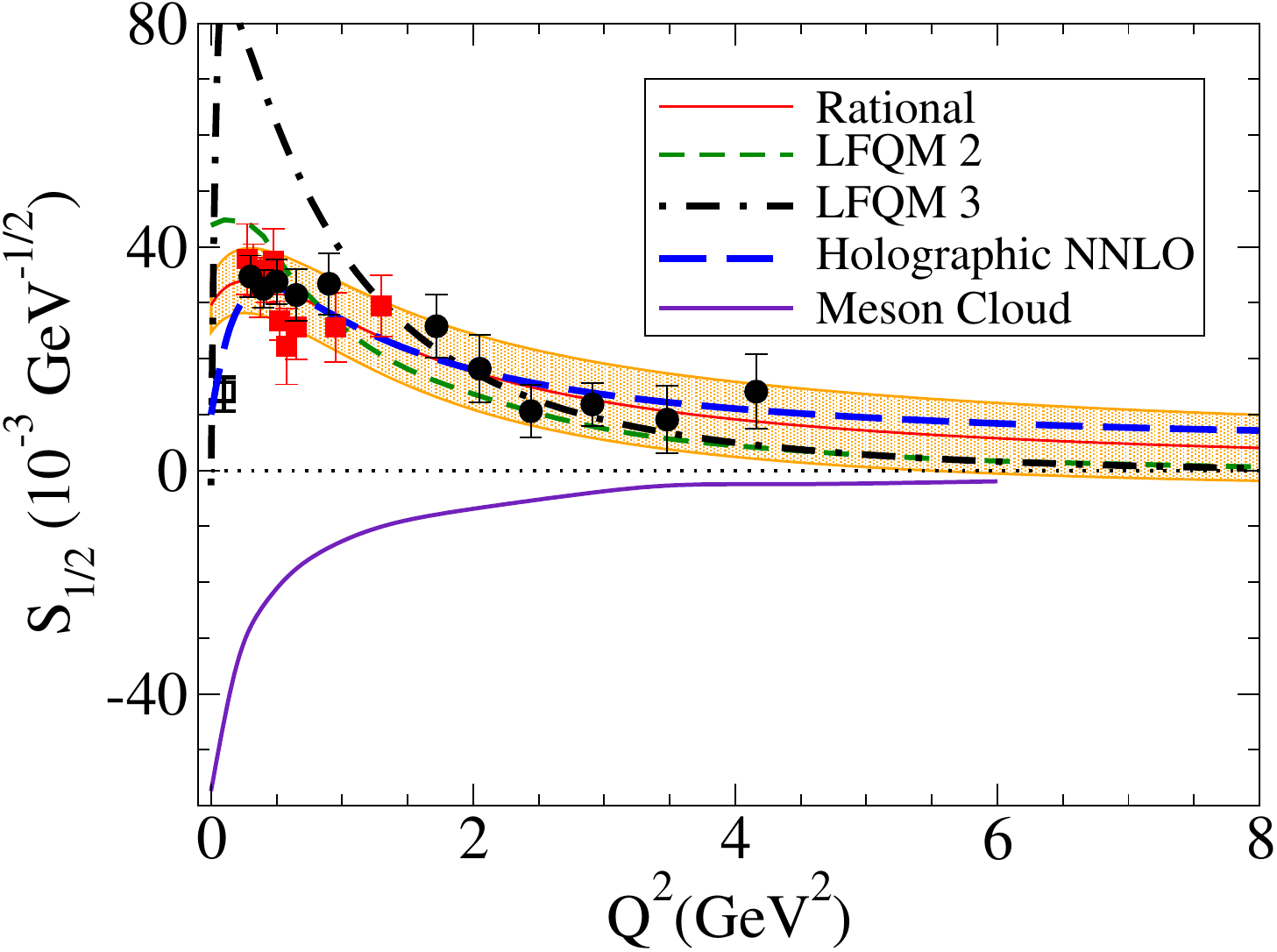}
    \caption{Left panel: Representative results of the recent dynamical coupled-channel (JBW) approach. Right panel, overview of other approaches (adapted from a recent review~\cite{Ramalho:2023hqd}), employing  strictly real transition form factors. The black and red data points show the extractions from the CLAS collaboration for single~\cite{CLAS:2009ces} and double~\cite{CLAS:2007bvs,Mokeev:2015lda} pion production, respectively.
    }
    \label{fig:TFF-results}
\end{figure}

Experimental observables are measured for real values of $W$. To access the transition form factors from the multipoles extracted in the procedure sketched above in a \emph{reaction independent} way, analyticity of both the multipoles as well as the scattering amplitude needs to be exploited for the (analytic) continuation to the resonance pole. Specifically, for any resonance (fixed total angular momentum $J$ and isospin $I$) at the resonance pole ($W_{\rm pole}$) one can write
\begin{equation}
    \mathcal{H}\propto\frac{\widetilde{\mathcal{H}}}{W-W_{\rm pole}}+\cdots\ ,
    \quad
    T\propto\frac{\widetilde{R}}{W-W_{\rm pole}}+\cdots\,,
\end{equation}
where $\mathcal{H}\in\{\mathcal{A_{1/2}},\mathcal{A_{3/2}},\mathcal{S_{1/2}}\}$ denotes a linear combination of the corresponding multipoles $\mathcal{M}$. Following  Ref.~\cite{Workman:2013rca} (see also Refs.~\cite{Ramalho:2023hqd, Ramalho:2025xim}), the transition form factors are then defined as 
\begin{equation}
\label{TFFdef}
	H(Q^2)=
	C_I\sqrt{\frac{p_{\pi N}}{\omega_0}\frac{2\pi(2J+1)W_{\rm pole}}{m_N\widetilde{R}}}\widetilde{\mathcal{H}}(Q^2)\,,
\end{equation}
where $H\in\{A_{1/2},A_{3/2},S_{1/2}\}$. Here, $\omega_0$ is the energy of the photon at $Q^2=0$, $m_N$ the nucleon mass, and the isospin factor~\cite{Drechsel:1998hk} $C_{1/2}=-\sqrt{3}$ and $C_{3/2}=\sqrt{2/3}$. Since at the pole the residues factorize into a product of incoming and outgoing effective couplings, the factor $\sqrt{\widetilde{R}}$ in the denominator removes the effective coupling (including its phase) connected to the outgoing final state from the expression, leaving the reaction independent information on the transition form factor.

\begin{figure}[t]
    \centering
    \includegraphics[width=0.49\linewidth]{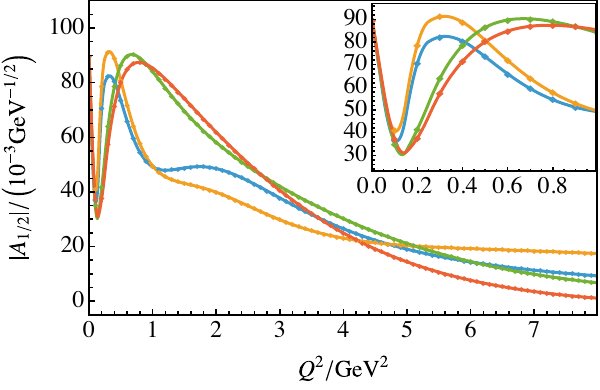}
    \includegraphics[width=0.49\linewidth]{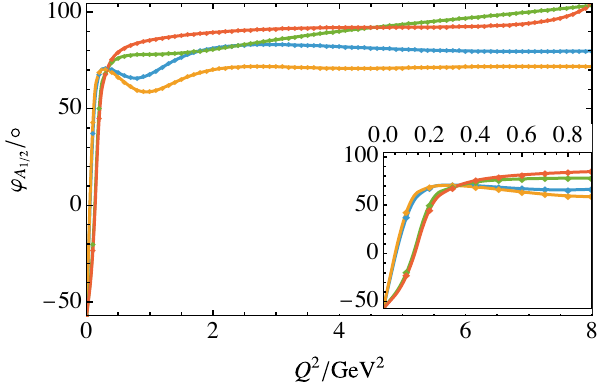}
    \\
    \includegraphics[width=0.49\linewidth]{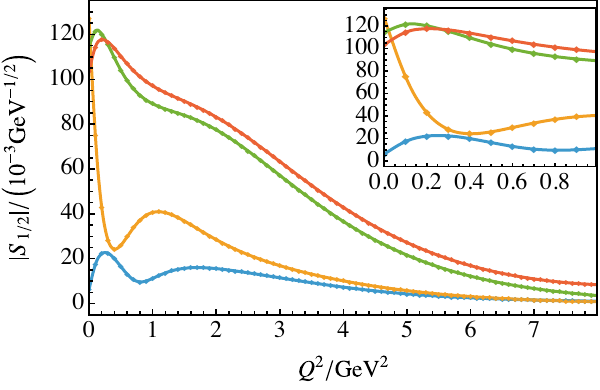}
    \includegraphics[width=0.49\linewidth]{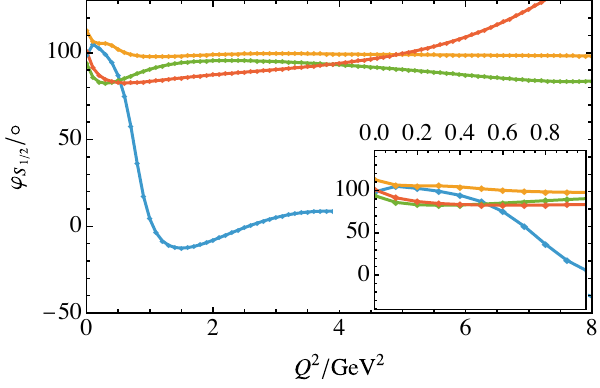}\\
    \caption{Results of the recent dynamical coupled-channel (JBW) approach on transition form factors in terms of their absolute values (left) and phases (right).}
    \label{fig:JBW-phases}
\end{figure}

Using the latest JBW solution which provides an adequate  description of the experimental data (e.g., $\chi^2_{\gamma^*}/{\rm dof}<1.5$) and the procedure outlined above one obtains the transition form factor of the Roper resonance~\cite{Wang:2024byt}.  In the left panel of Fig.~\ref{fig:TFF-results} we show four different solutions of the fits---for later use in Fig.~\ref{fig:JBW-phases} we show the corresponding absolute values and phases of the transition form factors. In the right panel of Fig.~\ref{fig:TFF-results} we show the results of other studies for the transition form factors as well as the TFFs provided by the CLAS collaboration---those should not be confused with data, since there is some procedure necessary to come from the measured data to the TFFs. Note that all the TFFs extracted using the  JBW approach describe the data equally well---although they appear to be rather different individually. Moreover, they all show non-negligible imaginary parts. In contrast to this,  the TFFs extracted by the CLAS collaboration are real valued. Furthermore, the comparison of the left and the right panels of Fig.~\ref{fig:TFF-results} suggests that the theoretical uncertainties assigned to the extraction by CLAS seem to be underestimated. 

The transition form factors provided by the CLAS collaboration are extracted from the multipoles using the assumption that the resonances are well represented by Breit-Wigner (BW) amplitudes with constant widths and real couplings to both the $\gamma N$ channel as well as the final states---for a comparison of the different extractions, see Ref.~\cite{Workman:2013rca}. A comparison of the left and right panels of Fig.~\ref{fig:TFF-results}, thus, suggests that not only are the theoretical uncertainties severely underestimated, but also the assumption that the Roper resonance can be described by a fixed mass BW function is not justified. This also puts into question the interpretations of CLAS TFFs for the Roper resonance provided in, e.g., Ref.~\cite{Burkert:2017djo}.

\section{Discussion and Summary}

We discuss the results shown in Fig.~\ref{fig:JBW-phases}. A first important observation is that the absolute value of none of the transition form factors shows a zero, contrary to those of Refs.~\cite{CLAS:2009ces, CLAS:2007bvs, Mokeev:2015lda}. Accordingly, an interpretation of that zero as a signature of the node in the Roper wave function that appears necessarily, if the resonance is a radial excitation of the nucleon, is hard to justify. However, what can be read off of Fig.~\ref{fig:JBW-phases} is that there is a very non-trivial hadronic dynamics taking place reflected in a strong $Q^2$-dependence of both the magnitude and the phase of the transition form factors. 

The J\"ulich-Bonn-Washington dynamical coupled-channel model does not have any explicit pole term for the Roper resonance which appears to be generated dynamically from the meson-exchange potential through the scattering equation~\cite{Krehl:1999km}. Given what is said above, this description appears to be consistent with the available data. This observation raises a reasonable doubt on the claim that the meson cloud effects are  negligible for the $A_{1/2}$ transition amplitude (see purple line in the upper right panel of Fig.~\ref{fig:TFF-results}). This is yet another important point questioning the interpretation of the CLAS TFFs provided in, e.g., Ref.~\cite{Burkert:2017djo}.

To summarize, transition form factors and couplings of resonances can be defined unambiguously only at the resonance poles. Here they typically develop non-trivial phases that should not be neglected. Applied to the case of the Roper resonance we demonstrated that:
\begin{itemize}
    \item[a)] From the presently available data the transition form factors, especially $S_{1/2}$, cannot be extracted with high accuracy.
    \item[b)] There is a very strong energy dependence in the phase especially of $A_{1/2}$ pointing at meson-baryon dynamics being very relevant for the structure of this resonance.
\end{itemize}
In particular, the latter observation challenges the validity of the simple interpretation of the Roper resonance as the first radial excitation at least on the basis of the current experimental situation. Improved data should allow for an increased extraction accuracy for the electro-magnetic multipoles in the future. Crucially, however, also a sophisticated theoretical tool-box -- based on a careful analytic continuation to the resonance poles -- is necessary to uncover universal resonance properties from these data.

\section*{Acknowledgements}
We are grateful to Gilberto Ramalho for very insightful discussions and  for providing the figures in the right panels of Fig.~\ref{fig:TFF-results}.
This work was funded in part by the Deutsche Forschungsgemeinschaft (DFG, German Research Foundation) as part of the CRC 1639 NuMeriQS–project no.~511713970, under Germany's Excellence Strategy - EXC 3107/1 - 533766364,
by the European Research Council (ERC) under the European Union's Horizon 2020 research and innovation programme (grant agreement No. 101018170),
by the MKW NRW under the funding code NW21-024-A,
and by the CAS President's International Fellowship Initiative (PIFI) under Grant Nos.~2025PD0022 (UGM) and 2025PD0087 (CH). 
The work of MM was further funded through the Heisenberg Programme by the Deutsche Forschungsgemeinschaft (DFG, German Research Foundation) – 532635001.
The numerical calculations were performed on JURECA-DC
of the J\"ulich Supercomputing Centre, J\"ulich, Germany.

\bibliographystyle{unsrt}
\bibliography{refs}
	
\end{document}